\documentclass[aps,prl,twocolumn,showpacs,floatfix]{revtex4}

\usepackage{amsmath,float,graphicx}

\begin{document}

\title{Sound propagation in elongated superfluid fermion clouds}

\author{P. Capuzzi}
\email{capuzzi@sns.it}
\author{P. Vignolo}
\email{vignolo@sns.it}
\author{F. Federici}
\email{fr.federici@sns.it}
\author{M. P. Tosi}
\email{tosim@sns.it}
\affiliation{NEST-INFM and
Classe di Scienze, Scuola Normale Superiore, I-56126 Pisa, Italy}

\begin{abstract}
We use hydrodynamic equations to study sound propagation in a
superfluid Fermi gas inside a strongly elongated cigar-shaped trap,
with main attention to the transition from the BCS to the unitary
regime. We treat first the role of the radial density profile in the
quasi-onedimensional limit and then evaluate numerically the effect of
the axial confinement in a configuration in which a hole is present 
in the gas density at the center of the trap.  We find that in a strongly
elongated trap the speed of sound in both the BCS and the unitary regime
differs by a factor $\sqrt{3/5}$ from that in a homogeneous
three-dimensional superfluid.  The predictions of the theory could be
tested by measurements of sound-wave propagation in a set-up such as
that exploited by M.R. Andrews {\it et al.}  [Phys. Rev. Lett.  {\bf
79}, 553 (1997)] for an atomic Bose-Einstein condensate.
\end{abstract}

\maketitle

%\section{\label{sec:intro}Introduction}

Strong evidence for a superfluid state in ultracold Fermi gases near
a Feshbach resonance has come from experimental studies using as
probes ballistic expansion
\cite{Ohara2002a,Bourdel2003a,Regal2003a,Zwierlein2003a,Jochim2003a},
collective modes \cite{Kinast2004a,Bartenstein2004a}, RF
spectroscopy \cite{Chin2004a}, and the generation of quantized
vortices \cite{Zwierlein2005a}.  Since the BCS weak-coupling limit
and the unitary strong-coupling limit are characterized by the same
collective-mode frequencies and expansion dynamics \cite{Hu2004a},
attention has been drawn to the study of sound-wave propagation as
allowing a clear identification of these two regimes. In particular Ho
\cite{Ho2004a} and Heiselberg \cite{Heiselberg2005a} have evaluated
the speed of first and second sound in a homogeneous Fermi
superfluid as functions of the coupling regime.  At low temperature
the first sound velocity is given by $u_1=v_{\rm
F}\sqrt{(1+(2/\pi)\hbar k_{\rm F}\,a)/3}\simeq v_{\rm F}/\sqrt{3}$
in the dilute BCS limit and by $u_1\simeq 0.37 v_{\rm F}$ in the
unitary limit. Here $v_{\rm F}$ and $k_{\rm F}$ are the Fermi
velocity and wavenumber, while $a$ is the $s$-wave scattering
length.

The main purpose of this Letter is to evaluate the propagation
of density perturbations in a superfluid Fermi gas as a function
of coupling strength in an experimentally realizable set-up.
This simulates the set-up used in the experiments of Andrews
{\it et al.} \cite{Andrews1997a} on the
propagation of sound pulses along the axis of an elongated cloud of
Bose-Einstein condensed $^{23}$Na atoms. In their
experiments a density perturbation is generated by turning on a laser at the
center of the trap and two pulses propagate
in opposite directions along the trap axis. We describe the
 dynamics of density fluctuations in the superfluid
by hydrodynamic equations, namely the continuity equation
\begin{equation}
\partial_t n+\nabla\cdot(n{\bf v})=0
\label{cont}
\end{equation}
and the Euler equation
\begin{equation}
m\partial_t {\bf v}+\nabla\left[\mu(n)+
V({\bf r})+\tfrac{1}{2}m{\bf v}^2\right]=0
 \label{eulero}
\end{equation}
for the time-dependent density profile $n({\bf r},t)$ and velocity
field ${\bf v}({\bf r},t)$ of the trapped gas. The equation of state
enters through the density-dependent chemical potential $\mu(n)$,
and $V({\bf r})$ describes the external potentials. The laser beam
is simulated in our study by a time-independent effective potential
$U(z)=U_0 e^{-z^2/w^2}$ having amplitude $U_0$ and width $w$,
which is turned on at the center of the trap.

{\it The Q1D configuration.}
We discuss first the quasi-onedimensional (Q1D) case corresponding to
$V({\bf r})=m\omega_\perp^2r_\perp^2/2$, where
$\omega_\perp$ is the radial trap frequency. This is amenable to
analytical treatment. The equilibrium density $n_0(r_\perp)$
is determined from the relation
$\mu(n_0)=\bar{\mu}-m\omega^2_\perp r^2_\perp/2$,
where $\bar{\mu}$ is determined by the normalization condition
$\int n_0(r_\perp) d^2r_\perp=\tilde{n}$, $\tilde{n}$ being the
linear density.
Linearization around equilibrium sets
$n({\bf r},t) = n_0+\delta n(r_\perp)e^{i(\omega t-qz)}$
and
$\mu(n)=\mu(n_0)+\left.\partial \mu/\partial n
\right|_{n=n_0}\!\!\!\delta n(r_\perp)e^{i(\omega t-qz)}.$
Thus Eqs. (\ref{cont}) and (\ref{eulero}) lead to the
eigenvalue equation
\begin{multline}
m\omega^2\delta n= q^2\left(n_0\left.{\partial \mu}/{\partial n}
\right|_{n=n_0}\!\!\!\delta n\right) \\ -\nabla_\perp\cdot\left[n_0\nabla_\perp
\left(\left.{\partial \mu}/{\partial n}
\right|_{n=n_0}\!\!\!\delta n\right)\right]
\label{collmode}
\end{multline}
and integration in the $(x,y)$ plane  yields
the dispersion relation
\begin{equation}
\omega = q\,\left(\frac{1}{m}{\int n_0\,{\partial \mu}/{\partial
  n}|_{n=n_0}\,\delta n\,d^2r_{\perp}}\Bigr/{\int \,\delta
  n \,d^2r_{\perp}}\right)^{1/2}.
\label{dispersion}
\end{equation}
The lowest collective mode of
the gas is obtained by taking $\delta n$ as the lowest-energy solution of Eq.\
(\ref{collmode}) with $q=0$, {\it i.e.}
$\delta n(r_\perp)\propto\left(\left.{\partial \mu}/
{\partial n}\right|_{n=n_0}
\right)^{-1}.$
Therefore the sound velocity is
\begin{equation}
u_1=\left(\frac{1}{m}\int n_0 d^2r_\perp \Bigr/\int
\left({\partial \mu}/{\partial n}|_{n=n_0}
\right)^{-1}d^2r_\perp\right)^{1/2}.
\label{soundgen}
\end{equation}
Equation (\ref{soundgen}) applies to all Q1D gases which can be
described by hydrodynamic equations.

For a power-law form of
the equation of state, $\mu(n)=\mathcal{C}n^{\gamma}$, the equilibrium
density profile takes the explicit form
$n_0(r_\perp)=\left({\bar\mu}/{\mathcal{C}}\right)^{1/\gamma}
\left(1-{r_\perp^2}/{R_{\rm TF}^2}\right)^{1/\gamma}$
where $R_{\rm TF}=\sqrt{2\bar\mu/m\omega_\perp^2}$ is the radial size
of the cloud.  Equation (\ref{soundgen}) then reduces to
\begin{equation}
u_1^2=
\frac{\gamma\bar\mu}{(1+\gamma)m}.
\label{eq:u1gamma}
\end{equation}
For $\gamma=1$ Eq. (\ref{eq:u1gamma}) gives the speed of sound pulses
propagating along the axis of a Q1D condensate of bosonic atoms
\cite{Kavoulakis1998a} or of molecular bosons formed by pairs of
fermionic atoms. For a Fermi gas in the BCS ($k_{\rm F}|a|\ll 1$) and the
unitary ($k_{\rm F}|a| \gg 1$) regime the equation of state is instead
proportional to the Fermi energy $\frac{\hbar^2}{2m}(3\pi^2n)^{2/3}$,
so that $\gamma=2/3$ and $u_1=\sqrt{2\bar\mu/5m}$.  In the former
limit at very low temperatures $\bar\mu=m v_{\rm F}^2/2$ and
$u_1=v_{\rm F}/\sqrt{5}$, while in the latter $\bar\mu=m
(1+\beta)v_{\rm F}^2/2$ with $\beta$ being a constant that has been
evaluated in various approaches \cite{Heiselberg2001a}. Quantum Monte
Carlo calculations find $\beta\simeq -0.57$
\cite{Carlson2003a,Astrakharchik2004} and then $u_1\simeq 0.29 v_{\rm
F}$.  These velocities thus differ by a factor $\sqrt{3/5}$ from those
in the homogeneous Fermi gas \cite{Heiselberg2005a} and would be
measured in a sound propagation experiment through the central portion
of a strongly elongated trap.  It is worth noticing that these results
differ from what a 1D calculation would predict: in the latter it is
straightforward to show that $u_1=\sqrt{\gamma\bar\mu/m}$, {\it i.e.}
$u_1=v_{\rm F}$ for a strictly 1D hydrodynamic Fermi gas.

\begin{figure}
\includegraphics[width=\columnwidth,clip=true]{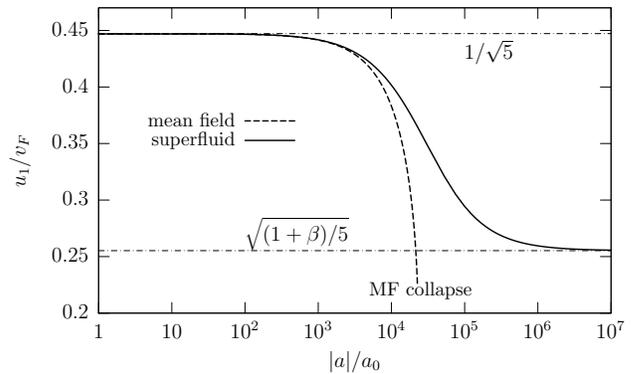}
\caption{Speed of sound $u_1$ in a Q1D Fermi gas (in units of the
Fermi velocity $v_{\rm F}$) as a function of the $s$-wave scattering
length $a$ (in units of the Bohr radius $a_0$ and on a log scale).
The full and the dashed line are the results from the equation of
state (\ref{eq:heis_mu}) and from the mean-field (MF) model,
respectively.  The dot-dashed lines mark the BCS (upper) and the unitary
(lower) limits.}
\label{fig1}
\end{figure}

At intermediate couplings, when $k_{\rm F}|a|$ is finite,
$\mathcal{C}$ depends on the density and several forms of the
equation of state have been proposed
\cite{Heiselberg2001a,Pieri2004a,Hu2005a}.  A possible solution is
to use again a power law with an effective exponent $\gamma$, which
may be inferred from a measured collective mode frequency
\cite{Hu2004a}.  The effective value of $\mathcal{C}$ can instead be
determined from the cloud size. In this work we focus on a gas with
a negative scattering length and choose the equation of state given
by Heiselberg in the Galitskii approximation \cite{Heiselberg2001a}.
This is
\begin{multline}
\mu(n)=\frac{\hbar^2}{2m}
(3\pi^2n)^{2/3}\,\\\times\left[1+\frac{c_1n^{1/3}}{1-c_2n^{1/3}}
\left(2+\frac{c_2n^{1/3}}{3(1-c_2n^{1/3})}\right)\right]
\label{eq:heis_mu},
\end{multline}
with $c_1=2a/(9\pi)^{1/3}$ giving the mean-field contribution
and $c_2=6[11-2\ln 2](3\pi^2)^{1/3}a/(35\pi)$ the second-order
contribution from the ladder diagrams, which takes into
account the opening of the superfluid gap. Equation 
(\ref{eq:heis_mu}) reproduces the result $\mu=
m[1 + 4 a k_{\rm F}/(3\pi)]\,v_{\rm F}^2/2$ in the limit of a dilute system, and
is approximately valid at intermediate densities in the unitary limit
where it gives $m(1+\beta)v_{\rm F}^2/2$. The ratio between $c_1$ and
$c_2$ in this case yields $\beta\simeq-0.67$. The results for the
sound velocity as obtained from a self-consistent
evaluation of the chemical potential and of the
equilibrium density are shown in Fig. \ref{fig1}. In
this calculation we have taken a linear density $\tilde{n}$ corresponding
to the chemical potential of  $2\times10^3$ fermions in a
cigar-shaped confinement with trap frequencies
$\omega_{\perp}=100\,{\rm s}^{-1}$ and $\omega_z = 6.3\,{\rm s}^{-1}$
(see below).  The BCS and unitary limits are
recovered for $a\rightarrow 0^-$ and $a\rightarrow -\infty$,
respectively.  The dashed line in Fig. \ref{fig1}
shows the sound velocity obtained within the
mean-field model for a dilute Fermi gas by setting
$c_2=0$ in Eq.\ (\ref{eq:heis_mu}).
The mean-field model well describes the BCS limit,
but as the coupling increases it predicts a strong
drop of the sound velocity heralding collapse.
The location and width of the transition from the BCS to the unitary
regime are determined by the system parameters.

\paragraph{Effect of axial potentials.}
A weak harmonic confinement along the $z$ direction induces a smooth
inhomogeneity in the axial density profile, which can be treated
within a local density approximation. Assuming again a power-law
equation of state, Eq.\ (\ref{soundgen}) leads to the velocity field
\begin{equation}
{v}_1(z) = \left(\frac{\bar\mu}{m}\frac{\gamma}{1+\gamma}\right)^{1/2}
\left(1-\frac{z^2}{Z_{\rm TF}^2}\right)^{1/2}, \label{eq:ldau1}
\end{equation}
with $Z_{\rm TF}=\sqrt{2\bar\mu/m\omega_z^2}$, which corresponds to
an oscillatory motion with frequency
$\omega=\omega_z/\sqrt{2+2/\gamma}$ and velocity $u_1$ at the center
of the trap. Even though $u_1$ depends on $\mathcal{C}$, the
oscillation frequency is the same in the BCS and in the unitary
limit, since the product between the wavenumber $q\propto 1/Z_{\rm
TF}\propto \sqrt{1/\bar\mu}$ and the velocity
$u_1\propto\sqrt{\bar\mu}$ does not depend on the chemical
potential.

\begin{figure}
\begin{tabular}{ccc}
\includegraphics[width=0.32\columnwidth,clip=true]{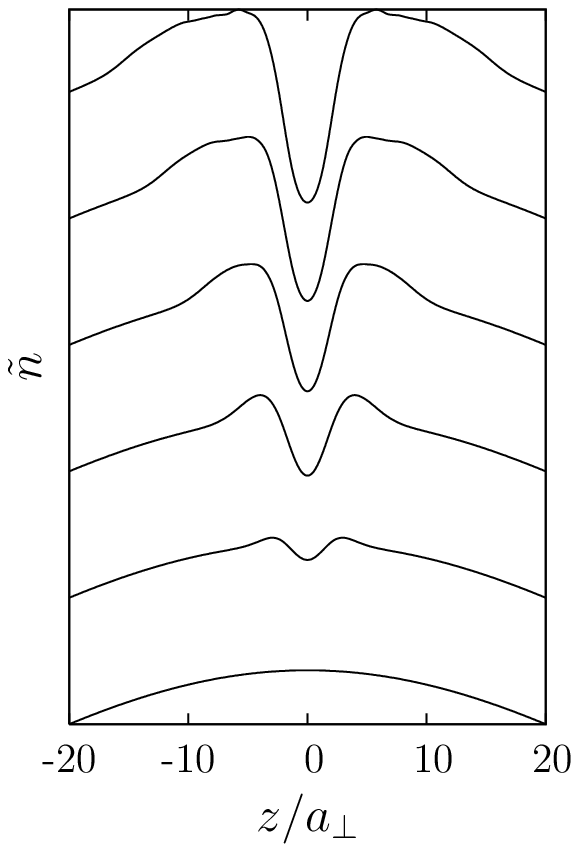}&
\includegraphics[width=0.32\columnwidth,clip=true]{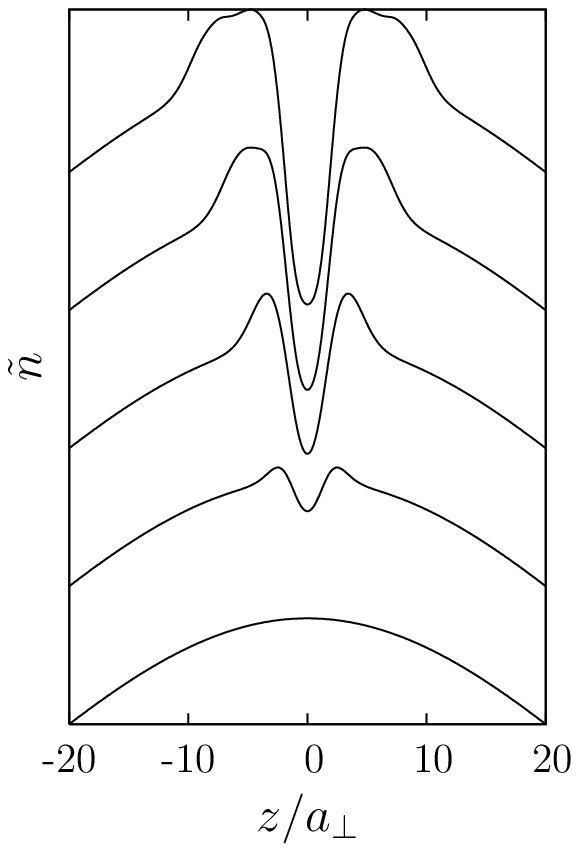}&
\includegraphics[width=0.32\columnwidth,clip=true]{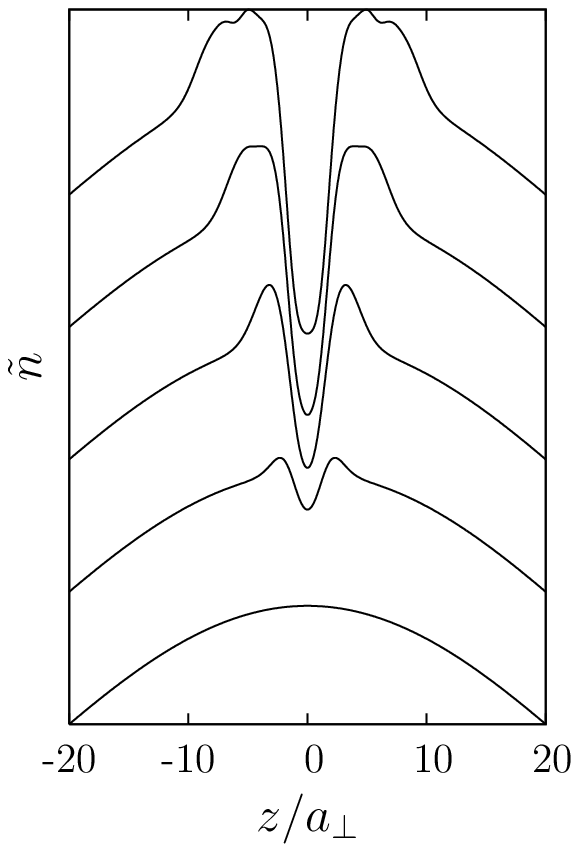}
\end{tabular}
\caption{\label{fig:integrated} Integrated density profile
$\tilde{n}$ (in arbitrary units) as a function of axial coordinate
$z$ (in units of $a_{\perp}=(\hbar/(m\omega_{\perp}))^{1/2}$) at
different times starting from the bottom at $t=0$ and increasingly
separated by $1.5/\omega_{\perp}$. The left, middle, and right panel
correspond to $a/a_0=-80, -10^5$, and $-10^6$. For the sake of
visibility the plotting range is smaller than the full extent of the
density profile.}
\end{figure}

We next carry out a numerical study of sound propagation using the
hydrodynamic equations (\ref{cont}) and (\ref{eulero}) in a set-up
simulating that used in the experiments of Andrews \textit{et al.}
\cite{Andrews1997a}. In addition to radial and axial confinements
provided by a strongly elongated harmonic trap, we switch on a
hole-digging potential $U(z)$ at the center of the trap. The atoms
are thereby expelled into two density perturbations that travel
towards the ends of the trap.  Since the perturbation velocity is
mainly in the $z$ direction and imaging techniques will provide
column densities, we focus on the evolution of the radially-integrated
density profile $\tilde{n}(z)$. The numerical calculations are
carried out for a system of $N=2\times 10^3$ $^{40}$K atoms equally
distributed in two spin-polarized states, with trap frequencies
$\omega_{\perp}=100\,{\rm s}^{-1}$ and $\omega_z = 6.3\,{\rm
s}^{-1}$. The potential $U(z)$ has a width $w$ equal to 5\% of the
axial width of the density profile and an amplitude $U_0$ of magnitude
up to $\hbar\omega_z$.  In Fig.\ \ref{fig:integrated} we illustrate
the time evolution of $\tilde{n}(z)$ for three values of the
scattering length, $a=-80, -10^5, \text{ and }-10^6$ Bohr radii. The
size of the cloud decreases on increasing the coupling strength and
at the same time the speed of the density perturbation also
decreases.

\begin{figure}
\includegraphics[width=\columnwidth,clip=true]{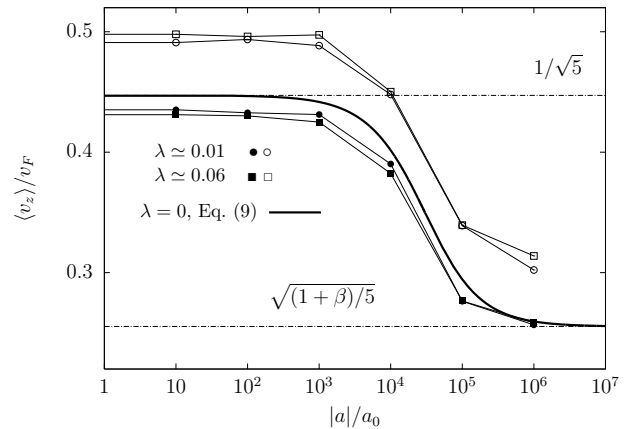}
\caption{\label{fig:velofin}Sound velocity in a superfluid Fermi gas
as a function of $|a|$ (in units of the Bohr radius $a_0$ and on a log
scale). Circles and squares correspond to the average speed of
perturbations propagating in a trap with $\lambda\simeq0.06$ and
$\lambda=0.01$. Empty and full symbols refer to $U_0/\hbar\omega_z=0.8$ and
$0.16$, respectively.  The thick solid line is from
Eq.~(\ref{soundgen}), and the dot-dashed lines mark the BCS (upper) and
unitary (lower) limits. The thin solid lines are guides to the
eye.}
\end{figure}

From the evolution of the integrated density we calculate the mean
velocity $\langle v_z\rangle$ of the perturbation in a time interval
$\Delta t = 5/\omega_{\perp}$. The results for $\langle v_z\rangle$ as
a function of the coupling strength and for two amplitudes of the
perturbing potential are shown in Fig.  \ref{fig:velofin}. The
comparison is also illustrated between the case with $\lambda\equiv
\omega_z/\omega_{\perp}\simeq 0.06$ and that with $\lambda=0.01$ at
the same zero-coupling Fermi velocity.  For weak perturbations (for
example $U_0=0.16\hbar\omega_z$, full symbols in Fig.\
\ref{fig:velofin}), giving $\delta n$ of the order of a few percent of
the equilibrium density at the trap center, the speed of sound for
both values of $\lambda$ is found to be in good agreement with the
theoretical prediction given in Eq.~(\ref{soundgen}) for the Q1D gas
(thick solid line). The slight difference seen in
Fig.~\ref{fig:velofin} may be accounted for by the
effect of the weak confinement causing the speed of sound to decrease
while the density perturbation moves away from the central region of
the trap (see Eq.~(\ref{eq:ldau1})).  Beyond the perturbative regime
(empty symbols in Fig.\ \ref{fig:velofin}, referring to $U_0=0.8\hbar\omega_z$
and $\delta n/n_0\sim 20\%$) we find an upward shift of the mean
velocity as a result of the increase in the local density.  The
overall behavior of $\langle v_z \rangle$ is still the same as in the
perturbative regime. The ratio between the values at zero and large
$|a|$ is about $0.61$, to be compared with $\sqrt{1+\beta}\simeq0.57$
for both the Q1D gas ($\lambda=0$) and the homogeneous gas.

\begin{figure}
\includegraphics[width=\columnwidth,clip=true]{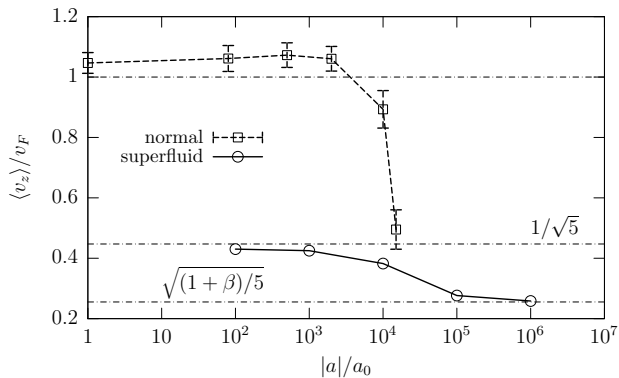}
\caption{\label{fig:vmedia}Sound velocity in an elongated trap as a
function of $|a|$ (in units of the Bohr radius $a_0$ and in a log
scale). Circles and squares are for a superfluid and for a normal
Fermi gas, respectively. The error bars for the normal-gas results are
from the standard deviation in the velocity distributions.
The dot-dashed lines show the zero-sound, the BCS, and the unitary
limit, from top to bottom.} 
\end{figure}

Finally, we compare in Fig. \ref{fig:vmedia} the sound velocity for
$\lambda\simeq 0.06$ in a superfluid Fermi gas with that in a normal
Fermi gas at a temperature $T=0.2\,T_{\rm F}$, $T_{\rm F}$ being the
Fermi temperature at zero coupling. The dynamics of the normal gas has
been studied by the Vlasov-Landau kinetic equations as described in
Ref.\ \cite{Akdeniz2003a}.  Our method involves binary collisions
between the particles and includes Pauli-blocking effects.
At low coupling the normal Fermi mixture is in the
collisionless zero-sound regime and sound propagation is associated
with an anisotropic deformation of the Fermi surface, the zero-sound
velocity being close to $v_{\rm F}$ as shown in Fig.
\ref{fig:vmedia}. With increasing attractions the velocity of the
perturbation drops.  This drop occurs at about $a\simeq -10^4 a_0$, and
can be attributed to both the transition towards the first-sound
velocity $v_{\rm F}/\sqrt{5}$ and the instability of the approaching
collapse of the normal state. The transition to the superfluid state
restores stability.

In summary, we have studied sound propagation in a superfluid Fermi
gas confined inside a strongly elongated harmonic trap. We have found
that the sound velocity diminishes as the superfluid goes from the BCS
to the unitary regime on increasing the strength of the attractions
between the atoms, and that in both limits the speed of sound differs
by a factor $\sqrt{3/5}$ from that in the superfluid three-dimensional
gas. The numerical study of the superfluid gas dynamics in a weak
axial confinement shows that the value of the speed of sound around in
the central region of the trap is close to that in the fully quasi-onedimensional
gas. However, in situations of experimental relevance, a density
perturbation of order 20\% could give rise to upward shifts of the
speed of sound by 10--12\%. These effects could be observed in
experimental set-ups of trapped superfluid Fermi gases and thereby
provide further insight in their superfluid state.

This work was partially supported by an Advanced Research
Initiative of Scuola Normale Superiore di Pisa.

\end{document}